\documentclass[10pt, conference]{IEEEtran}
\IEEEoverridecommandlockouts

\ifCLASSINFOpdf
\else
\fi
\usepackage{seqsplit}
\usepackage{xspace}
\usepackage{SIunits}            
\usepackage{courier}            
\usepackage[scaled]{helvet} 
\usepackage{url}                  
\usepackage{listings}          
\usepackage{enumerate}
\usepackage[colorlinks=true,allcolors=blue,breaklinks,draft=false,bookmarks=false]{hyperref}
\usepackage[dvipsnames,table,xcdraw]{xcolor}
\usepackage{caption}
\usepackage{tikz}

\usepackage{amsmath,amstext,amsthm}
\usepackage{thmtools}
\usepackage{thm-restate}
\usepackage{tabularx}
\usepackage{balance}

\usepackage{commands}

\begin{document}

\newcommand{\twosigmadisclaimer}{The views expressed herein are not necessarily the views of Two Sigma Investments, LP or any of its affiliates (collectively, ``Two Sigma''). The information presented herein is only for informational and educational purposes and is not an offer to sell or the solicitation of an offer to buy any securities or other instruments. Additionally, the information is not intended to provide, and should not be relied upon for investment, accounting, legal or tax advice. Two Sigma makes no representations, express or implied, regarding the accuracy or completeness of this information, and you accept all risks in relying on the above information for any purpose whatsoever.}
%
\title{Who you gonna call? \\ Analyzing Web Requests in Android Applications}




%

\author{%
\IEEEauthorblockN{%
Marianna Rapoport\IEEEauthorrefmark{1},
Philippe Suter\IEEEauthorrefmark{2}\IEEEauthorrefmark{3}\thanks{\IEEEauthorrefmark{3}\twosigmadisclaimer},
Erik Wittern\IEEEauthorrefmark{2},
Ond\v{r}ej Lh\'{o}tak\IEEEauthorrefmark{1} and
Julian Dolby\IEEEauthorrefmark{2}%
}
\IEEEauthorblockA{%
\IEEEauthorrefmark{1}University of Waterloo, Waterloo, Canada\\
\IEEEauthorrefmark{2}IBM Research, Yorktown Heights, NY, USA\\
\IEEEauthorrefmark{3}Two Sigma Investments, LP, New York, NY, USA\\
mrapoport@uwaterloo.ca,
philippe.suter@twosigma.com,
witternj@us.ibm.com,
olhotak@uwaterloo.ca,
dolby@us.ibm.com%
}}


\maketitle

\begin{abstract}
Relying on ubiquitous Internet connectivity, applications on mobile devices frequently perform web requests during their execution. They fetch data for users to interact with, invoke remote functionalities, or send user-generated content or meta-data.
These requests collectively reveal common practices of mobile application development, like what external services are used and how, and they point to possible negative effects like security and privacy violations, or impacts on battery life.
In this paper, we assess different ways to analyze what web requests Android applications make.
We start by presenting dynamic data collected from running $20$ randomly selected Android applications and observing their network activity. Next, we present a static analysis tool, \stringoid, that analyzes string concatenations in Android applications to estimate constructed URL strings. Using \stringoid, we extract URLs from $30,000$ Android applications, and compare the performance with a simpler constant extraction analysis. Finally, we present a discussion of the advantages and limitations of dynamic and static analyses when extracting URLs, as we compare the data extracted by \stringoid from the same $20$ applications with the dynamically collected data.
\end{abstract}


%
\IEEEpeerreviewmaketitle


\section{Introduction}
\label{sec:introduction}
Many mobile applications use the ubiquitous Internet connection of mobile devices to make \emph{web requests}, meaning network requests based on web technologies like HTTP as a transfer protocol, or JSON or XML as data formats.
For example, using web requests, applications consume third party services through \emph{web APIs} to interact with otherwise inaccessible resources, like data or functionalities. Reliance on such services is especially important as mobile devices have limited storage and processing capabilities. The largest public catalog of web APIs, ProgrammableWeb~\cite{ProgrammableWeb}, documents the rising number of such APIs, listing currently over $16,500$ APIs, up from only $100$ in 2005 and $5,000$ in 2012.
Mobile applications also perform web requests to obtain data from or transmit data to proprietary backends, including sending user-generated data, data about the usage of applications, or the context in which they are used.

Knowing what web requests are made by mobile applications is desirable for multiple reasons.
Regarding individual applications, knowing about web requests is important for security and privacy reasons. Users care about privacy and security issues related to mobile devices in particular~\cite{Chin:2012}, and their concerns to this regard have been intensely studied~\cite{Fernandez:2013}. Web requests do not have to be disclosed to users, who have little or no insight or even control over what requests are performed, and thus over what data is sent, when, and to whom.
Apart from privacy concerns, web requests may also eat into users' data plans and reduce battery life -- in fact, a recent study finds that mobile applications' network usage is the leading factor in the energy consumption of mobile devices~\cite{Li:2014}.
Across applications, information about web requests performed from mobile applications helps to understand the usage of web APIs.
API directories and recommendation services such as API Harmony~\cite{Wittern:2016} inform developers on which APIs are commonly used by applications, both individually and in combinations with each other.
From analyzing source code, they could learn which specific API endpoints are frequently used, and how (e.g., what query parameters are used; what data is sent in request bodies).

Information on the requests mobile application make and when, while valuable, is typically not made available to users or developers. The resulting question is how can we learn what web requests mobile applications make?

In this paper, we address this question with three contributions:
\begin{itemize}
  \item We present a dataset of dynamically monitored HTTP(S) traffic from $20$ randomly selected Android applications. Our analysis of that data provides insights into application behavior and also shows the difficulties to obtain web request-data dynamically with high coverage.
  \item We present a publicly available tool called \stringoid to automatically scan Android applications for string concatenation operations indicative of the construction of URLs, and therefore of potential web requests.
  \item We perform two experiments to statically extract URLs from Android applications, first using a simple constant string extraction and second relying on the presented \stringoid tool. We present results from performing these experiments on the $30,000$ most used Android applications obtained from the Playdrone dataset~\cite{viennot2014measurement}.
\end{itemize}
Our comparison of the statically and dynamically obtained data highlights the challenges of our work, motivating a discussion on the limitations of dynamic and static analyses in general.

In the remainder of this paper, we start by outlining preliminaries on web requests and establish some metrics which we use in our experiments in Section~\ref{sec:preliminaries}.
We present our method for dynamically collecting request data from $20$ Android applications and an analysis of that data in Section~\ref{sec:dynamic}. Next, we shift our focus to statically learning about web requests in Section~\ref{sec:static}. We describe our publicly available \stringoid tool to statically extract URL strings from Android application source code in Section~\ref{sec:static_tool}. We present an analysis of the data resulting from applying a simple constant string extraction and \stringoid to $30,000$ Android applications in Section~\ref{sec:static_data}. Finally, we compare and discuss the dynamic and static analysis approaches in Section~\ref{sec:discussion}, before presenting related work in Section~\ref{sec:related} and concluding in Section~\ref{sec:conclusion}.


\section{Preliminaries}
\label{sec:preliminaries}

In this section, we introduce concepts and terminology used in the rest of the paper, and give an overview of the problems involved in determining the web requests Android applications make.

Our work focuses on a specific class of outgoing network requests made by applications, namely HTTP requests. While applications can in principle make requests over other protocols, we believe that HTTP is by far the most common one. Such requests can be to third-party APIs (e.g. ad networks, authentication providers), or to services known only to the application developers (e.g. server backend supporting the application functionality).

An HTTP request is characterized by several components: the URL, the method (e.g. GET, POST), the headers, and the body. A corresponding response from the server has another set of headers and a body.
In this paper, we concentrate on the URL part of requests, as it contains the information relevant to answering the question of which services or hosts are accessed.

\subsection{Anatomy of URLs}
\label{sec:preliminaries_urls}

By design, URLs are constructed as hierarchical sets of information.
Consider for instance the URL:

\tinycodeblock{http://example.com/api/users/SMcDuck?extended=true}

In the example above, we call \code{http} the \emph{protocol}, \code{example.com} the \emph{domain name}, and \code{/api/users/SMcDuck} the \emph{path}.
We call \code{extended} a \emph{query parameter key} (referred to as \emph{key}), \code{true} a \emph{query parameter value} (referred to simply as \emph{value}), and \code{extended=true} a \emph{query parameter key-value pair} (referred to simply as \emph{key-value pair}).


The parts that form a URL are of different nature:
Within an application, some parts are typically \emph{static}, for example the \emph{host} and \emph{domain} of a request.
Other parts are \emph{dynamic}, for example the above path parameter used to identify a specific user or the query parameter value.
Dynamic URL parts are typically provided at runtime by either an application user or the environment in which an application runs in.

When dynamically observing requests, URLs are always fully-formed, however, for the purpose of static analysis, it is useful to define a concept of \emph{URL pattern}.
Patterns make explicit the parts which can be filled in by the application.
In the following example, the dynamic parts of a set of URLs are denoted with brackets (\emph{placeholders}):

\tinycodeblock{http://example.com/api/users/[ ]?extended=[ ]}

Note that a pattern may have no placeholders, so all URLs are also patterns.

The different parts of a URL or URL pattern contain different information regarding a request. From the fully-formed example URL above, we can learn that data is transferred without being encrypted as the HTTP protocol is used (rather than HTTPS). We also learn from the domain and first path component that the request targeted the ``api'' hosted by ``example.com''. Furthermore, a ``users'' resource was invoked, which was identified by ``SMcDuck'' and ``extended'' data was requested. When inferring URL patterns from multiple observed URLs, it might be learned that users are identified by alphanumeric words of a certain length or that the query parameter ``extended'' expects Boolean values. Overall, URL act as a proxy that allows to infer a considerable amount of information on web requests.

\subsection{Quantifying URL sets}
\label{sec:preliminaries_components}
When running dynamic and static analyses designed to collect sets of URLs, 
we are interested in making quantitative and qualitative assessments of the
results. We could count the number
of URLs retrieved for a given application, but there are two main reasons for which
this approach is unsatisfactory.

First, analyses can produce large sets of very similar URLs; in a dynamic experiment, this can happen if an application makes repeated requests, changing only a small set of parameters. A static analysis tries to enumerate all possible URLs created by a given code fragment, and if that fragment contains conditionals, the set can in principle be very large.

Second, as we have seen above, the information represented in URLs is hierarchical. A set of URLs differing only in the value of a query parameter, for instance, is qualitatively different from a set of the same size showing many different top-level domains.

To address both issues, when applicable, we break down URLs into components: domains, paths, query parameter keys and query parameter values.
This separation can also be applied to \emph{patterns} of URLs.
Consider for instance the three patterns below, two of which contain a placeholder:

\vspace{1mm}
\tinycode{http://example.com/api/info?user=bar\&limit=12}

\tinycode{http://example.com/[ ]?show=[ ]}

\tinycode{http://example.com/api/list?sort=[ ]}
\vspace{1mm}

\noindent From this set, we compute the following elements:
\begin{itemize}
    \item one domain name: \tinycode{example.com}
    \item three paths: \tinycode{/api/info}, \tinycode{[ ]}, and \tinycode{list}
    \item four keys: \tinycode{user}, \tinycode{limit}, \tinycode{show}, and \tinycode{sort}
    \item four values: \tinycode{user=bar}, \tinycode{limit=12}, \tinycode{show=[ ]}, \tinycode{sort=[ ]}
\end{itemize}
As a convention we count placeholders as one distinct value. Finally, so as to not conflate, e.g., identical paths appearing in URLs for different domains, we organize sets of URLs into:
\begin{enumerate}
    \item a set of domains (\emph{domains})
    \item a set of pairs of a domain and a path (\emph{path-pairs})
    \item a set of triples of a domain, a path, and a query parameter key (\emph{key-triples})
    \item a set of quadruples of a domain, a path, a query parameter key, and a query parameter value (\emph{value-tuples})
\end{enumerate}
This methodology properly accounts for URLs that contain a subset of information of another, or where the query parameters are re-ordered.

%
%
%
%
%
%
%
%
%
%
%
%
%


\section{Dynamic Analysis}
\label{sec:dynamic}
The first approach to obtain data about the web requests a mobile application makes is to observe the network traffic that application makes. This approach, ideally, provides not only an exact account of what requests are actually made, but also when and in what frequency while using the application.

\subsection{Data collection method}
\label{sec:dynamic_method}
To obtain dynamic web request data, we ran $20$ randomly selected applications from the Playdrone data set~\cite{viennot2014measurement} in the Android emulator and monitored the outgoing and incoming HTTP traffic.

\subsubsection{Runtime environment}
As a virtual device, we used a Google Nexus 4 running Android $4.2$ (codenamed ``Jelly Bean'', API level 17). To collect the possibly encrypted HTTP traffic, we proxied the traffic of the Android emulator through the \code{mitmproxy} tool\footnote{\code{mitmproxy} 0.14; \url{http://mitmproxy.org/}} by configuring the network settings of the virtual device. The proxy acts as a man-in-the-middle: after installing the appropriate SSL certificate on the virtual device, the proxy listens in on the requests made by the device, whether they are encrypted or not, forwards them to their destination, listens to the responses and re-encrypts them before serving them to the device. We were thus able to monitor and log the full content of all HTTPS requests and responses made from the virtual device. To actually collect the desired data, we wrote an inline script used by the proxy to log information on every HTTP and HTTPS request, including the request's URL, HTTP method, resulting HTTP status code, and the received data.

\subsubsection{Application selection}
In order to eliminate any bias in the selection of applications, we picked them
randomly from the set of the $5,000$ most popular applications in the Playdrone
dataset. For a variety of reasons, we could not simply keep the first random picks:
\begin{itemize}
    \item $16$ applications could not be installed through the Android Debug Bridge (ADB), reportedly due to container errors
    \item $32$ applications could be installed, but crashed immediately upon opening
\end{itemize}
We further dismissed $16$ applications that we considered unfit for the experiment:
\begin{itemize}
    \item $6$ applications had limited or no functionality, for example just setting a single, fixed, background image
    \item $4$ applications were designed to perform arbitrary user-defined HTTP requests, such as browsers or network tools
    \item $2$ applications made no HTTP request
    \item $2$ applications were excluded because we could not properly use them as they were in a language we do not understand
    \item $2$ applications were written in JavaScript in a Java container, and thus unfit for analysis by \stringoid
\end{itemize}
The above reasons mean that we ended up attempting to install and test
$84$ applications to build our test suite of $20$ applications.

\subsubsection{Testing protocol}
For each of the $20$ applications, we used the following protocol:
we installed the application on the virtual device and uninstalled all other applications to avoid them making requests in the background.
We then started and manually used the installed application for $5$ minutes.
When using the application, we attempted to cover as many features as possible.
If an action in the application activated another application (typically, opening a browser), we immediately returned back to the application in focus.
If necessary, we created user accounts to be able to fully use the application.
After $5$ minutes, we closed the application and stopped recording the traffic.

\subsection{Dynamic request data}
\label{sec:dynamic_data}
Overall, using the above data collection method, we observed $5,975$ requests performed by the $20$ applications. Request counts per application of our experiment are shown in Table~\ref{table:dynamic}. The dynamically collected request and response data is publicly available \cite{zenodo_dataset}.

\subsubsection{Data summarization}
The column ``overall'' in Table~\ref{table:dynamic} shows for each application the number of HTTP requests it made in the $5$ minute window, and the column ``unique'' shows the corresponding number of distinct URLs being invoked.
What is immediately striking is that the lowest number of observed HTTP requests is $81$, meaning that every application assessed made at least on average one request every $3.5$ seconds. Half of the applications made more than $215$ requests (around one request every $1.4$ seconds) and one quarter of the applications made more than $324$ requests (more than one request per second). The most active application made $1,059$ requests in the $5$ minute interval, meaning it performed on average $3.53$ requests per second.
These numbers are significant even without knowing about the nature of requests as network activity has been found to be the major impact of mobile applications on the battery life of mobile devices~\cite{Li:2014}.

\begin{figure}
  \includegraphics[width=\columnwidth]{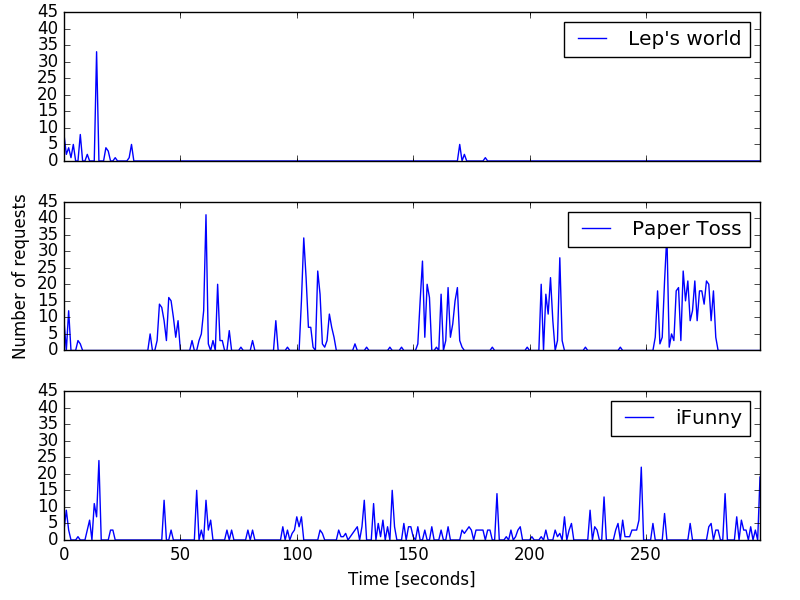}
  \caption{Web requests per second for selected applications.}
  \label{fig:requests_over_time}
\end{figure}

Figure~\ref{fig:requests_over_time} shows three selected examples of when requests occurred over the five minute experiment interval. In ``Lep's world'', a jump-and-run game, most requests are performed shortly after starting the application and entering a game, at which point advertisements are shown to the user. In contrast, the game ``Paper toss'' (which performed the most requests in the experiment) occasionally presents users with advertisements, leading to related bursts of requests. The final example, ``iFunny'', shows comedic pictures such as memes to users, which are loaded on-demand while interacting with the application.

Across all $5,975$ requests, $5,277$ ($88.3\%$) used \texttt{GET}, and the remaining $698$ ($11.7\%$) used \texttt{POST}. No other HTTP methods were observed.
Only $2,986$ requests ($50\%$) succeeded, meaning they resulted in a status code between \texttt{200} and \texttt{399}. Of the remaining $2,989$ requests, only $71$ ($2.4\%$) reported an HTTP status code, while $1,460$ requests ($48.8\%$) aborted due to a client error, and $1,446$ requests ($48.3\%$) did not succeed because the server disconnected before the request could succeed. Client disconnects mean that the client aborted the TCP connection underlying the HTTP request before the server managed to respond, for example due to timeouts. One possible reason for server disconnects is that the requested service is no longer available, which could be due to the relatively old age of the tested applications. Despite the relatively high degree of unsuccessful requests, there were no noticeable interferences with the functionalities of the test applications, raising questions on the exact nature of the observed requests.


\subsubsection{Nature of observed requests}
In this section, we aim to assess the nature of the dynamically observed requests.
One way to do so is to assess the content-type of their responses. Of the $2,986$ requests resulting in a response with an HTTP status code, $2,863$ define a \texttt{content-type} header. The remaining $123$ responses simply did not contain such a header. Table~\ref{table:content_types} provides an overview of content-type categories. To create the categories, different content-types were selected using the regular expressions reported in the table. The chosen categories allow to classify the content types in a mutually exclusive and collectively exhaustive way.
As the table shows, media files (images, video, and audio) make out the most common type of responses, namely $43.4\%$. 
Next, source code (HTML, JavaScript, and CSS) accounts for $37.3\%$ of responses,
followed by data (JSON, XML, and plain text) with $12.1\%$.

\begin{table}
\begin{tabularx}{\columnwidth}{X|l|r|r}
\textbf{Content type} & \textbf{Used regex} & \textbf{\# requests} & \textbf{\# ad requests} \\
\hline
Image      & $^\wedge$\texttt{image}                  &   1,198 & 412  (34.4\%) \\
HTML       & \texttt{html}                            &     616 & 266  (43.2\%) \\
JavaScript & \texttt{javascript}                      &     372 & 108  (29.0\%) \\
JSON       & \texttt{json}                            &     266 &  35  (13.2\%) \\
Stream     & \texttt{octet|stream}                    &      83 &   6   (7.2\%) \\
CSS        & \texttt{css}                             &      79 &  27  (34.2\%) \\
Cache      & \texttt{cache}                           &      65 &  63  (96.9\%) \\
Xml        & \tiny{\texttt{(application|text)/.*xml}} &      59 &   6  (10.2\%) \\
Video      & $^\wedge$\texttt{video}                  &      35 &   3   (8.6\%) \\
Font       & \texttt{font|ttf}                        &      31 &   2   (6.5\%) \\
Zip        & \texttt{zip}                             &      27 &   1   (3.7\%) \\
Plain text & \texttt{plain}                           &      21 &   3  (14.3\%) \\
Audio      & $^\wedge$\texttt{audio}                  &      10 &   1  (10.0\%) \\
Thrift     & \texttt{thrift}                          &       1 &   0   (0.0\%) \\ \hline
\textbf{Sum} &                                        & $3,077$ & $933$         \\ \hline \hline
\end{tabularx}
\caption{Categories of \texttt{content-type} of responses received overall and from advertisement services. Percentage denote the share of advertisement responses among all responses for each content-type.}
\label{table:content_types}
\end{table}


While response content-types help understand the nature of a request, they do not help to answer the question why a request was made. To answer this question, we aimed to obtain a qualitative overview of what types of requests are made. We looked at the request URLs and find that overall $293$ different domains are targeted by the requests. $3,029$ requests (a little over $50\%$), however, target only one of $23$ domains.
We manually assessed these domains and find that
$8$ of them (accounting for $1,430$ requests, or $23.9\%$ of all requests) target advertisement services,
$6$ of them (accounting for $689$ requests, or $11.5\%$ of all requests) target content delivery networks, and
$3$ of them (accounting for $250$ requests, or $4.2\%$ of all requests) target APIs. All three categories were determined by manual investigation into the domains in question. The remaining requests out of the manually assessed sample targeted websites that do not lend themselves for categorization.







Motivated by the manual assessment, we aim to focus on request to advertisement services. We make use of publicly available lists of known advertisement services~\cite{Adaway:2017,Fanboy:2017}. These lists are typically used for ad-blocking software, allowing to tag requests based on their URL to either target advertisement services or not. We ran all observed URLs against the latest versions of these lists as of January 25th 2017 and thus found that $1,826$ or $30.6\%$ of all request target advertisement services, of which $933$ ($51\%$ of ad requests; $30.3\%$ of successful requests) succeeded.
We assessed the content-types of succeeding ad requests in column ``\# ad requests'' in Table~\ref{table:content_types}. The presented percentage number in that column indicates how many of all requests for a certain content type were caused by ad requests. From the table we can see that requests targeting advertisement services are over-proportionally responsible for requests to media files (images, video, and audio) and source code (HTML, JavaScript, and CSS), accounting for $33.5\%$ and $37.6\%$ of them respectively (thus exceeding their overall share of requests of $30.6\%$). On the other hand, they are relatively under-represented when it comes to requests for data (JSON, XML, and plain text), accounting for only $12.1\%$ of them.



\subsection{Discussion and takeaways}
\label{sec:dynamic_discussion}
Dynamic request data, on first sight, seems to be extremely precise: we observe the actual requests a mobile applications makes. However, precision can be impacted by requests made by other applications on a device or background processes. Trying to associate requests to their origin application is a recent research problem~\cite{Yao:2015,Tongaonkar:2013}. In addition, observed requests do not necessarily represent \emph{all} requests an application would make if its usage was extended, thus impeding recall. Even though we tried to use applications extensively in our experiment, we were not necessarily able to cover all functionalities that may induce requests. This is because sometimes parts of an applications are only accessible with membership accounts or to paid members, functionalities may only be available on certain devices, at certain times, or in certain regions, or because functionalities may depend on other apps being installed.

Despite these limitations, the dynamic analysis revealed large amounts of requests for every one of the tested $20$ applications. Assessing these requests revealed that a significant amount of them, namely over $30\%$, target advertisement services. In future work, it would be interesting to see whether these results generalize to larger and more diverse sets of applications (e.g. paid applications).



\section{Static Analyses}
\label{sec:static}
While the dynamic analysis presented in Section~\ref{sec:dynamic} reflects the actual requests made by an application, it also has some downsides. Performing the dynamic analysis requires a lot of effort to install the application and manually execute it, even if the required virtual device and proxy are already set up. In addition, as discussed in Section~\ref{sec:dynamic_discussion}, the manual execution may not cover the whole application and thus leave potential requests undetected. To scale the analysis of requests, another option is to apply static analysis on the application code. In this section, we present a static analysis tool called \stringoid built to extract URLs, reflecting web requests, from Android applications. We furthermore perform a large-scale experiment with \stringoid on $30,000$ Android applications and compare its results with basic constant extraction.


\subsection{The \stringoid tool for static analysis}
\label{sec:static_tool}

\stringoid is a static analysis tool developed for our inquiry into web requests made by Android applications. It is designed to be fully automatic, and with the goal of achieving relatively low running times (seconds, or a small number of minutes) on common applications, so that it can be applied to a large set of applications. \stringoid is publicly available.\footnote{\url{https://github.com/amaurremi/stringoid}}

\begin{figure}
\begin{lstlisting}
String getWeatherUrl() {
    String host = "https://weather.example.com";
    String timeQuery = "time=";
    if (forToday) timeQuery += "today";
    else timeQuery += this.time;
    String locationQuery = "city=" + getCity();
    return String.format(
        "%s?%s&%s", host, timeQuery, locationQuery);
}
\end{lstlisting}
\caption{Example of code creating URLs in Java.}
\label{fig:weather-code}
\end{figure}

\stringoid takes as input an Android application archive (\code{.apk} file) and produces a \emph{set of string patterns} representing URLs.
As an illustration of the type of information \stringoid can extract from programs, consider the method in Figure~\ref{fig:weather-code}.
The method can build URLs following two patterns:

\tinycodeblock{https://weather.example.com?time=today\&city=[ ]}

\noindent and

\tinycodeblock{https://weather.example.com?time=[ ]\&city=[ ]}

...where the placeholders possibly indicate the results of \code{getCity()} and \code{this.time}. The purpose of \stringoid is to automatically build such representations of URL patterns from applications compiled from code such as the above.

The tool operates in four main steps, which we describe below. Note that we only present superficially the static analysis techniques in the interest of readability.

\subsubsection{Reading the program structure}

The first step is to ingest the Android application and to construct an in-memory representation of its code structure.
For this task, \stringoid uses WALA \cite{fink2012wala}, a toolkit for static analysis of Java and Android applications. WALA constructs, for each method of each class (including libraries bundled with the application), a representation of the control-flow graph. Note that in our setting, WALA operates on bytecode.

\subsubsection{Estimating \code{StringBuilder} pointer sets}

The second step determines the correspondence between program variables and
instances of \code{StringBuilder}. This step is necessary largely because of
the difference between Java source code and Java bytecode: while the
representations are notoriously close, the most important difference affecting
\stringoid is that string concatenations (\code{a + b} in Java) are rewritten by the compiler
into \code{.append} operations on \code{StringBuilder} instances. Reasoning about objects is generally difficult in static analyses because of aliasing (the relationship between variables and the objects they can point to at each program point). The second step tracks aliasing to \code{StringBuilder} objects, in effect allowing the next step to reason as if the Java compiler had not transformed the code.

\subsubsection{Constructing automata}

\begin{figure}
\centering
\begin{tikzpicture}[->,>=stealth',auto,semithick]
  \node[initial,state,inner sep=2pt, minimum size=0pt] (A) {$q_0$};
  \node[state,inner sep=2pt, minimum size=0pt] (B) [right = 3\dist of A] {$q_1$};
  \node[state,inner sep=2pt, minimum size=0pt] (C) [right = 1.2\dist of B] {$q_2$};
  \node[state,inner sep=2pt, minimum size=0pt] (D) [right = 2.2\dist of C] {$q_3$};
  \node[state,inner sep=2pt, minimum size=0pt] (E) [below = 2\dist of D] {$q_4$};
  \node[state,inner sep=2pt, minimum size=0pt] (F) [left = 1.3\dist of E] {$q_5$};
  \node[state,inner sep=2pt, minimum size=0pt] (G) [left = 2\dist of F] {$q_6$};
  \node[state,accepting,inner sep=2pt, minimum size=0pt] (H) [left = 3\dist of G] {$q_7$};

  \path (A) edge  node {\codesf{``https://...''}} (B)
        (B) edge  node {\codesf{``?''}} (C)
        (C) edge  node {\codesf{``time=''}} (D)
        (D) edge  [bend left] node {\codesf{``today''}} (E)
            edge  [bend right] node [left] {\codesf{\textbf{this}.time}} (E)
        (E) edge node {\codesf{``\&''}} (F)
        (F) edge node {\codesf{``city=''}} (G)
        (G) edge node {\codesf{getCity()}} (H);
\end{tikzpicture}

\caption{Automaton representing the set of URLs constructed in Figure~\ref{fig:weather-code}.}
\label{fig:weather-automaton}
\end{figure}
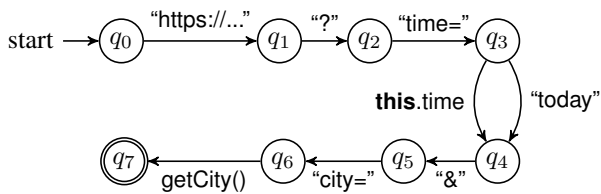

The third step uses the information obtained in the previous one, and analyzes each concatenation operation on \code{StringBuilder}s, and each use of \code{String.format} in the program. From that information and from the logical structure of the methods (if-branches and loop-conditions), it constructs \emph{automata} representing the sets of strings possibly constructed in the program. As a simplifying measure and to guarantee performance proportional to the size of the applications, \stringoid traverses loops at most once. We found this tradeoff to be acceptable based on the assumptions that in practice few URLs would be constructed in loops. A direct consequence of this design decision is that all automata produced by \stringoid are \emph{acyclic}.
Figure~\ref{fig:weather-automaton} shows the automaton corresponding to the code shown in Figure~\ref{fig:weather-code}.
We chose automata for their compact representation of many possible code paths.

\subsubsection{Outputing results}

The final step filters out automata that definitely do not represent URLs; for instance, if they start with a constant string that is clearly not a URL prefix. Finally, \stringoid outputs a serialized version of all remaining automata, which can be consumed by post-processing tools to compute statistics, or e.g. to enumerate URL patterns represented. In our experiments described below, for instance, we typically only keep URL patterns that can be processed by Python \code{urllib} library.
Figure~\ref{fig:weather-json} shows this representation for the automaton in Figure~\ref{fig:weather-automaton}.

\begin{figure}
\begin{lstlisting}
{ "url" : {
    "concat" : [
      { "kind" : "constant",
        "value" : "https://weather.example.com" },
      { "kind" : "constant", "value" : "?" },
      { "kind" : "constant", "value" : "time=" },
      { "kind" : "field",
        "type" : "java.lang.Date",
        "name" : "time" },
      { "kind" : "constant", "value" : "&city=" },
      { "kind" : "method-call",
        "return-type" : "java.lang.String",
        "name" : "getCity()" } } ] },
  "methods" : [ "java.lang.String getWeatherUrl()" ] }
\end{lstlisting}
\caption{Serialized automaton representing the URL patterns from Figure~\ref{fig:weather-code}.}
\label{fig:weather-json}
\end{figure}

\subsection{Limitations}
\label{sec:stringoid_limitations}

Every static analysis is built around a set of tradeoffs, and \stringoid is no exception. As already mentioned, \stringoid will not detect URL patterns that depend on more than one traversal of a loop in their construction. In addition, because the analysis is intra-procedural (i.e. does not consider execution across multiple methods), patterns may be less precise than desired. The pointer set construction phase may also introduce imprecision in programs that manipulate \code{StringBuilder} instances explicitly (as opposed to those generated by the Java compiler), if the aliasing relations are difficult to track (for instance, if a \code{StringBuilder} is temporarily stored in a variable).

In addition to the sources of false negatives listed above, \stringoid can also produce false positives. The main reason is the \emph{path-insensitive} nature of the analysis; for instance, if a method contains two \code{if}-branches with the same condition, \stringoid will consider four possible executions, while in reality only two are possible (both conditions are true or both are false).

Despite these limitations, we have found \stringoid in practice to be able to produce large sets of URL patterns and, through sampling, have not observed instances of obvious false positives.

\subsection{Data collection methods}
\label{sec:static_method}

Using \stringoid, we are able to apply the analysis of web requests to a large number of applications. One source for these applications is the Playdrone dataset~\cite{viennot2014measurement}. Playdrone is a distributed mining effort to collect binaries as well as meta data for over a million of Android applications from the Google Play store.

For our analysis, we made use of the most recent snapshot of the dataset (from October 2014), which is made publicly available on the Internet Archive.\footnote{\url{https://archive.org/details/android_apps}} From that dataset we selected the $30,000$ most popular applications, based on the number of downloads made available as part of application meta data, for which the application binary (\code{.apk} file) is also publicly available (this excludes paid applications). The raw data of these $30,000$ APK files plus metadata totaled in approximately $330$~GB.

To extract URLs from this input data, and get a sense of the capabilities of \stringoid at scale, we performed two experiments:
First, we ran a simple constant extraction analysis:
it looks for constant strings in the disassembled source code that look like URLs, i.e. that start with the (case-insensitive) prefix \code{http://} or \code{https://}. We implemented this analysis as part of \stringoid for convenience, although similar results can be obtained with the Android disassembler (\code{dexdump}) and \code{grep}. An added benefit of implementing the analysis on top of the same toolchain as \stringoid is that it lets us estimate the overhead of the full analysis (see below).
Second, we used the full automaton-building analysis of \stringoid to extract URL patterns.

\subsection{Static request data}
\label{sec:static_data}
\label{sec:static_data_comparison}

We ran both experiments on all $30,000$ applications using Apache Spark\footnote{\url{http://spark.apache.org/}} on a cluster of four 8-core machines. The total CPU time spent in the analyses was $99.5$ and $122.2$ hours for the constants and \stringoid analyses respectively. This lets us estimate the average overhead of \stringoid as $22.8\%$, or $2.8$ seconds per application on average. For $945$ applications ($3.2\%$), at least one analysis failed to produce a result, due to errors beyond our control in the Android bytecode processing library which \stringoid relies on. Whenever one analysis failed to produce a result, we excluded the results for both. (In the following we still refer to $30,000$ applications for presentation clarity.) All produced datasets are publicly available \cite{zenodo_dataset}.
Table~\ref{table:largescaleaggregate} shows the size of the sets of URL components extracted by each analysis on the $30,000$ applications, expressed using the methodology outlined in Section~\ref{sec:preliminaries_components}.
Note that by definition, the results produced by \stringoid are strict supersets of those produced by the constants analysis, because each constant string will at least be represented by an automaton with a single transition, even if it is never concatenated to anything else.

\begin{table}
\begin{tabularx}{\columnwidth}{X|rrrr}
                        &   Domains &   Path-pairs &     Key-triples & Value-tuples \\ \hline
     Constants          &  $47,920$ &  $142,791$ &   $35,440$ &  $55,211$ \\
 {\stringoid}           &  $50,483$ &  $207,047$ &   $83,014$ & $170,981$ \\ \hline
Improvement &  +5.3\% & +45.0\% & +134.2\% & +209.7\%
\end{tabularx}
\caption{Comparison of \stringoid with constant string extraction on $30,000$ applications.}
\label{table:largescaleaggregate}
\end{table}

We can see from the results that the improvements get higher as we move to elements more ``to the right'' of URLs. This is explained by the fact that in practice URLs are constructed left-to-right. The more concatenations \stringoid can detect, the deeper the information about the URL. Recovering such a wealth of information for a large set of applications also lets us compute some statistics that can reveal trends in mobile applications, and detect cases of complex URL creations in the applications.

\subsubsection{Macro-perspective on web requests}
\label{sec:static_data_macro}
Analyzing request data from large amounts of applications allows to derive insights into the usage of external services.

\begin{figure}
  \includegraphics[width=\columnwidth]{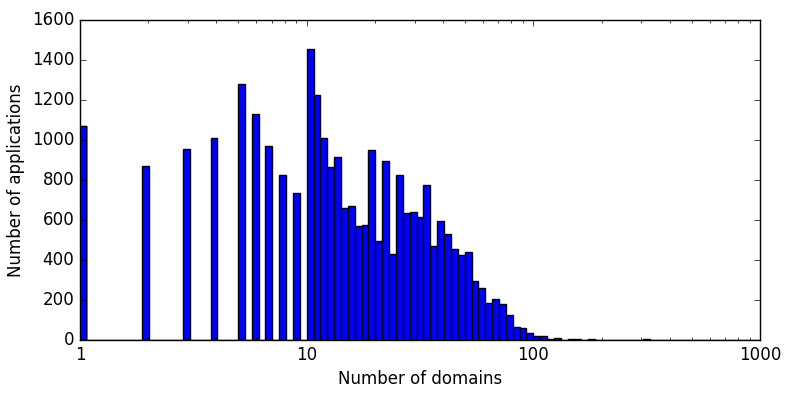}
  \caption{Histogram of unique domains per application. Note that the $x$-axis is log-scaled.}
  \label{fig:domains_per_app}
\end{figure}

One question we can attempt to answer with the data computed by \stringoid is the variety of Internet resources accessed by applications. Figure~\ref{fig:domains_per_app} shows a histogram of the number of unique domains per application. While the average number of unique URLs found per application is $19.9$, roughly $62.5\%$ of applications are found to contain URLs with that many or fewer unique domains. For half the  applications, $14$ or fewer domains were found. On the other hand, a comparatively small fraction of applications include considerably more unique domains in their URLs, for example, only $22$ applications reveal URLs with $150$ or more different domains.  The highest number of domains per application is an outlier with value $4,206$ (not displayed on the graph for readability, the second highest has 534 unique domains); it is the application ``World Newspapers'' which advertises itself as a media directory and which contains links to thousands of news organizations around the world. Another notable anomaly, although barely visible in Figure~\ref{fig:domains_per_app}, are $5$ applications in which we find URLs with the same $308$ domains each. These are related translation applications by the same publisher, which share a common code base.

\begin{figure}
  \includegraphics[width=\columnwidth]{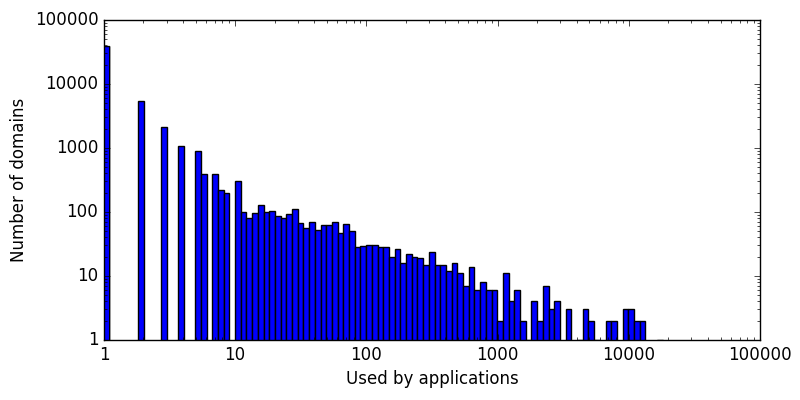}
  \caption{Histogram of application usage per domain. Note that both axes are log-scaled.}
  \label{fig:apps_per_domain}
\end{figure}

A related question is the relative importance of Internet resources; with the data from \stringoid, we can phrase it as computing the distributions of the number of applications that use a domain, which is the dual to the previous question. Figure~\ref{fig:apps_per_domain} illustrates this distribution. On average $10.6$ applications denote URLs for a given domain, and $38,466$ (or $76.2\%$) of these domains are found in URLs of a single application only. The relatively high mean and standard deviation of $221.6$ are explained by a comparatively small fraction of domains that are invoked by many applications. These findings point to a power-law distribution of the usage of domains by mobile applications. We expect this to come at least in part from the standard pattern of exposing a private service to one or a select few dedicated applications. This hypothesis is additionally supported by the subset of domains which represent hard-coded IP addresses, which we can assume are used by developers who have control over the servers. Overall, we found $1,703$ such IP address across all applications, $1,314$ of which were only found in a single application.

At the other end of the power distribution, our results highlight the importance of a few selected resources. Table~\ref{table:top_10} shows the $10$ most commonly used domains across all $30,000$ applications. Notably but perhaps not surprisingly on the Android platform, \emph{all} of them relate to Google offerings (AdMob is also a Google company).

\begin{table}
  \begin{tabular}{l|l|r}
    \textbf{No} & \textbf{Domain}    & \textbf{Number of apps.} \\ \hline
    1  & media.admob.com             & 16,725 \\
    2  & plus.google.com             & 13,158 \\
    3  & www.google.com              & 12,776 \\
    4  & googleads.g.doubleclick.net & 11,602 \\
    5  & www.googleapis.com          & 11,105 \\
    6  & www.google-analytics.com    & 10,693 \\
    7  & ssl.google-analytics.com    & 10,671 \\
    8  & play.google.com             & 9,986 \\
    9  & www.googletagmanager.com    & 9,139 \\
    10 & market.android.com          & 7,818 \\
  \end{tabular}
\caption{\label{table:top_10}Top 10 domains used across most applications. We removed \code{schemas.android.com}, \code{hostname}, and \code{details} from this list as these domains do not translate to actual HTTP requests.}
\end{table}

\subsubsection{Selected results}
\label{sec:static_data_selected}
\begin{figure*}
  \includegraphics[width=\linewidth]{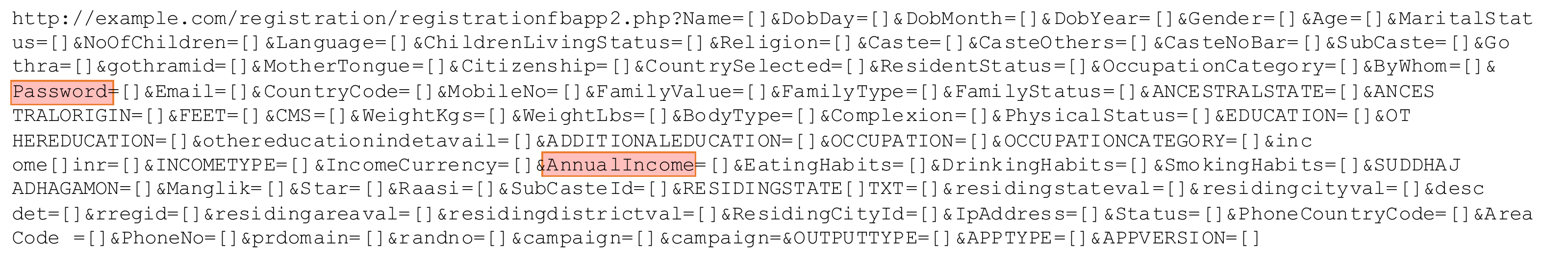}
  \caption{One of the longest patterns computed by \stringoid, exposing a lack of security in a dating application transmitting sensitive private information -- highlights are marked. The original domain name was redacted.}
  \label{fig:indiandating}
\end{figure*}

Beyond aggregate data, the URLs produced by \stringoid further provide means to detect interesting patterns on an application level. For example, Figure~\ref{fig:indiandating} shows one of the longest patterns computed by \stringoid, from a dating application.

The pattern corresponds to an HTTP GET request that we concluded the application makes at the end of the registration phase. It is a perfect example of a very long request where all keys are present in the application and thus exposed by \stringoid, and all values will be filled in at runtime. What is staggering is the amount of personal and highly sensitive information that is encoded in the request (salary, religious beliefs, etc.). While a user of a dating application will certainly realize that this information will be communicated to the server (since he or she needs to provide it), \stringoid exposes that it is most likely sent over the network without any encryption.

Applications can expose users' secrets through URLs, but they occasionally also expose the secrets of their developers. One URL pattern retrieved by \stringoid has the following structure (some keys and values are omitted for simplicity):

\vspace{1mm}
\tinycode{https://ex.com/search?q=[~]...\&clientsecret=OSJ...}
\vspace{1mm}

\noindent ...where the secret key is a long private chain of characters. In this example, in reconstructing fragments of the URL, \stringoid assembled various program variables and was able to replace the secret key. Note that the secret appears after a placeholder, meaning that the simpler constant extraction would not be enough to correctly place it in the URL.
Looking through the complete set of results, we found $2,420$ unique instances of key-value pairs where the key contained either the word ``secret'' or ``key''. We expect that a researcher targeting access credentials for specific services would have no trouble recovering many from \stringoid results.

\subsection{Discussion and takeaways}
\label{sec:static_discussion}

We have demonstrated that our dedicated static analysis designed for extracting URL patterns outperforms simple constant extraction, and is applicable at scale. The results are quantitatively better, extracting more of each type of URL component, and particularly elements such as query parameter keys and values.

Because URL patterns are longer and richer, the results are also qualitatively better, unlocking the potential for detecting e.g. complex web API invocations. Our cursory macro analyses show that popular services can be identified from looking at sufficiently many applications.
We expect that the publication of our datasets will enable further analyses at scale, such as security leak detections.


\section{Comparing and Contrasting Analyses}
\label{sec:discussion}
We have looked at analyzing web requests from mobile applications using either dynamic or static techniques. A subsequent question is: how do the results compare to another? And where do they individually shine?


\begin{table*}
  \begin{tabularx}{\textwidth}{X|ccc|ccc|ccc|ccc}
  \textbf{Application} 
  & \multicolumn{3}{c}{\textbf{Domains}}
  & \multicolumn{3}{c}{\begin{minipage}{0.9in}\centering \textbf{Path pairs}\end{minipage}}
  & \multicolumn{3}{c}{\begin{minipage}{0.8in}\centering \textbf{Key triples}\end{minipage}}
  & \multicolumn{3}{c}{\begin{minipage}{0.8in}\centering \textbf{Value tuples}\end{minipage}} \\
    & $\mid D\mid$ & $\mid D \cap S \mid$ & $\mid S\mid$ & $\mid D\mid$ & $\mid D \cap S \mid$ & $\mid S\mid$ & $\mid D\mid$ & $\mid D \cap S \mid$ & $\mid S\mid$ & $\mid D\mid$ & $\mid D \cap S \mid$ & $\mid S\mid$ \\
  \hline
 Lep's World                 &   3  &   5  & 18     &      13  &   3  &  39     &      80 &  0 & 25      &    107 & 0 &  25 \\
 Xperia Z                    &  13  &   3  &  8     &      20  &   2  &  19     &      86 &  1 & 24      &    124 & 1 &  95 \\
 MP3 Cutter and Ringtone ... &   3  &   2  &  3     &      16  &   2  &   5     &      84 &  0 &  1      &    288 & 0 &   1 \\
 Paper Toss                  &  99  &   3  & 17     &     195  &   3  &  22     &     566 &  0 & 44      &  1,013 & 0 &  50 \\
 Fotor                       &   8  &   7  & 13     &      59  &   3  &  37     &      54 &  0 & 19      &    245 & 0 &  19 \\
 HangMan Game                &  15  &   2  &  5     &      69  &   1  &  12     &      55 &  0 &  1      &     58 & 0 &   2 \\
 GO Weather EX               &   8  &   7  & 43     &      15  &   7  &  76     &      87 &  9 & 44      &     97 & 4 &  64 \\
 Pocket                      &  20  &   4  & 20     &      59  &  10  &  89     &      91 &  0 & 35      &    106 & 0 &  41 \\
 Where is that?              &  22  &   4  & 55     &      54  &   1  &  72     &     320 &  0 & 48      &    434 & 0 &  49 \\
 Bubble Blast                &   5  &   2  & 42     &      20  &   0  &  69     &      60 &  0 & 85      &    235 & 0 & 106 \\
 Quick PDF Scanner           &   5  &   5  & 10     &      64  &   4  &  14     &     118 &  0 &  4      &    259 & 0 &   5 \\
 Nearby Messenger            &   2  &   4  & 34     &       7  &   6  &  41     &      50 &  0 & 27      &     79 & 0 &  27 \\
 Tapatalk                    &  17  &  11  & 56     &      93  &  19  & 156     &     144 &  3 & 24      &    179 & 0 &  29 \\
 Rage Meme Camera            &   5  &   4  &  6     &      29  &   0  &  12     &      81 &  0 & 23      &    181 & 0 &  23 \\
 Retrica                     &   2  &   2  & 18     &      12  &   2  &  57     &      28 &  0 &  2      &     29 & 0 &   3 \\
 Logo Quiz Ultimate          &   6  &   7  & 71     &      22  &   6  & 964     &     210 &  0 & 41      &    395 & 0 &  59 \\
 BeyondPod                   &  22  &   3  & 25     &      31  &   0  &  52     &      41 &  0 & 14      &     45 & 0 &  15 \\
 Broken Screen               &   5  &   5  & 14     &      27  &   3  &  25     &     118 &  0 &  1      &    157 & 0 &   1 \\
 iFunny                      &  16  &   5  & 56     &      76  &   0  &  80     &     417 &  0 & 38      &    489 & 0 &  39 \\
 Wisielec: Kto zostanie ...  &   8  &   0  &  1     &      32  &   0  &   1     &      41 &  0 &  0      &     70 & 0 &   0 \\
 \hline
 \textbf{Total}              & 284  &  85  & 515    &     913  &  72  & 1,842    &    2,731 & 13 & 500     &  4,590 & 5 & 653 \\ \hline \hline
  \end{tabularx}
  \caption{Comparison of found URL components (cf. Section~\ref{sec:preliminaries_components}). $\mid D\mid$ = found uniquely in dynamic analysis; $\mid D \cap S\mid$ = found in dynamic \emph{and} static analysis; $\mid S\mid$ = found uniquely in static analysis.}
  \label{table:comparison}
\end{table*}

Table~\ref{table:comparison} compares URLs, broken down into different components ``domains'', ``path pairs'', ``key triples'', and ``value tuples'' (as defined in Section~\ref{sec:preliminaries}) that we either observed solely in the dynamic analysis, solely in the static analysis using \stringoid, or in the dynamic \emph{and} in the static analysis.
In the first columns, we see that already a significant amount of domains appears only either in the static or dynamic data.
This tendency increases when looking at the path pairs, key triples, and value tuples respectively.
In the remainder of this section, we attempt to explain and illustrate these observations.

\subsection{URLs found only in dynamic data}
\label{sec:discussion_dynamic}
One explanation for URL components only present in dynamic data is that they are never to be found in the statically analyzed code to begin with. Table~\ref{table:dynamic} presents the overall and unique counts of URLs we collected for each application in the $5$ minute experiments. However, many of the URLs of observed requests do not originate from the application code.
On the one hand, URLs may stem from application resources other than the application code, (for instance, in a separate \code{.xml} file).
Furthermore, URLs may stem from the content received from previous requests. As a simple illustration, consider the very common pattern of displaying an advertisement in an application: the application first makes a request to an advertisement broker, the broker returns the relevant information, including the URL of an image to display. The device then makes a request to obtain that image and finally display it. In the example described above, the first requested URL (the broker) originates from the application code, but the second one (the image) originates from the response received to the first request. Even a perfect static analysis has no way of recovering the information from the second URL, as it is external to the application.
Finally, URLs may depend on specific user actions or input. Such information is likely to be used, for example, as values in query strings, which helps explain the significantly higher number of observed value tuples in the dynamic data.


\newcommand*\rot{\rotatebox{90}}

\begin{table}
\begin{tabularx}{\columnwidth}{l|cc|cc}
& \multicolumn{2}{c}{\begin{minipage}{1in}\centering \textbf{URLs observed dynamically}\end{minipage}}
& \multicolumn{2}{c}{\begin{minipage}{1.2in}\centering \textbf{URLs matched outside of code}\end{minipage}}
\\
  \textbf{Application Name}
& \centering overall
& \centering unique
& \begin{minipage}{0.4in}\centering in static\\resources\end{minipage}
& \begin{minipage}{0.4in}\centering in\\responses\end{minipage}
\\ \hline
{\footnotesize Lep's World}         &      85 &      22 &      - &    14 \\
{\footnotesize Xperia Z}            &      86 &      28 &      1 &    26 \\
{\footnotesize MP3 Cutter...}       &     156 &      63 &      - &    63 \\
{\footnotesize Paper Toss}          &   1,059 &     324 &      - &   323 \\
{\footnotesize Fotor}               &     324 &     124 &      - &    44 \\
{\footnotesize HangMan Game}        &     224 &      73 &      - &    72 \\
{\footnotesize GO Weather EX}       &     193 &      27 &      - &    16 \\
{\footnotesize Pocket}              &     299 &      86 &      - &    75 \\
{\footnotesize Where is that?}      &     320 &      92 &      - &    85 \\
{\footnotesize Bubble Blast}        &     171 &      54 &      - &    53 \\
{\footnotesize Quick PDF Scanner}   &     326 &     118 &      - &    87 \\
{\footnotesize Nearby Messenger}    &     147 &      18 &      - &    11 \\
{\footnotesize Tapatalk}            &     834 &     145 &      - &   111 \\
{\footnotesize Rage Meme Cam...}    &     303 &      73 &      - &    48 \\
{\footnotesize Retrica}             &      91 &      15 &      - &    11 \\
{\footnotesize Logo Quiz Ultimate}  &     207 &      59 &      - &    41 \\
{\footnotesize BeyondPod}           &     183 &      33 &     13 &    18 \\
{\footnotesize Broken Screen}       &     335 &      44 &      - &    24 \\
{\footnotesize iFunny}              &     551 &     114 &      - &    83 \\
{\footnotesize Wisielec: Kto...}    &      81 &      37 &      - &    36 \\
\hline
\textbf{Total}                      &   5,975 &   1,549 &     14 & 1,241 \\ \hline \hline
\end{tabularx}
\caption{Classification of monitored application requests and comparison with static analysis results.}
\label{table:dynamic}
\end{table}

The columns ``URLs matched outside code'' in Table~\ref{table:dynamic} indicate how many (unique) URLs were matched to static resources in the application packages and the contents of responses to other requests. To compute these sets, we first ran an automatic string matcher. It identified all URLs in resources and $417$ URLs in request responses. Exact matching is not sufficiently precise, however; many responses contain for instance HTML or JavaScript content, which itself includes relative links or dynamically constructs URL fragments. We further manually inspected all remaining URLs with a custom-built tool highlighting which fragments were found in responses. When in doubt, we checked the contents of the responses to establish a probable cause if possible. We erred on the side of classifying a URL as originating from the application code. Using this labor-intensive procedure, we manually classified $825$ URLs as originating from responses and $294$ as originating from application code.

This effort shows that in practice, a very large fraction of dynamically observed requests can potentially be explained by factors independent of the application code. To consider an extreme case, in the application ``Paper Toss'', out of $324$ unique request URLs, $323$ were found in responses.

\subsection{URLs found only in static data}
\label{sec:discussion_static}
On the other hand, we observe that a significant fraction of URLs is only detected by the static analysis.

One reason for this observation is that the dynamic analysis may simply fail to produce desired results. As discussed in Section~\ref{sec:dynamic_discussion}, one challenge of the dynamic analysis is how to ensure coverage of an application. In some cases, thus, the static analysis produces valid URLs, for which, however, we fail to trigger the corresponding request when executing the application in our experiments.
As a concrete example, in the application \emph{Pocket}, \stringoid identified URLs which we could manually attribute to $28$ API endpoints for the domain \code{getpocket.com}, and which we were not able to trigger by exercising the application (\code{getpocket.com/v3/acctchange}, \code{getpocket.com/v3/listinfoupdate}, etc.). These URLs not only seem valid, but also support a motivated use case to learn about services or APIs from analyzing web requests.


Another source of URLs reported only in the static analysis is dead code in applications: developers often include libraries which provide functionality beyond what is required in the application. For instance, a library for facilitating authentication may contain URLs pointing to multiple providers even if the application is configured to use only one. Because \stringoid analyzes all methods regardless of reachability from application entry points, it will report URLs for all providers. In consequence, the static analysis finds URLs including \code{https://api.twitter.com/oauth}, \code{https://m.facebook.com/login.php}, and \code{https://www.googleapis.com/auth} across various applications.

Finally, like all practical static analyses, \stringoid is not fully precise and may overapproximate the set of concatenations that can be produced in valid executions. While this is somewhat mitigated by the fact that we only keep URL (patterns) that can be processed with Python \code{urllib}, there are no formal guarantees on the precision of the result set, as discussed in Section~\ref{sec:stringoid_limitations}.


\section{Related Work}
\label{sec:related}
Related work has addressed the problem of understanding the network activities of mobile applications through dynamic analyses.
A set of works assesses mobile application network traffic for security and privacy reasons. For example, it attempts to detect malware that infects Android applications based on observing applications' network traffic patterns, relying on features like request sizes, frequencies, and IP addresses~\cite{Shabtai:2014}. Or, a tool called Securacy, installed on mobile devices, monitors IP addresses and ports invoked from the device, and checks whether secure connections (HTTPS) are used to identify potential security violations~\cite{Ferreira:2015}.
Other works attempt to associate monitored requests with the mobile applications from which they originate, relying for example on HTTP headers~\cite{Yao:2015} or on previously learned network patterns of in-app advertisement services used by applications~\cite{Tongaonkar:2013}.
All of these works are either limited to un-encrypted traffic, or rely on information that can be collected in any case, like IP addresses. In contrast, in our dynamic analysis, we used a proxy server with a man-in-the-middle facility to obtain information about the URLs being invoked even when traffic is encrypted.



On the static analysis aspect, beyond extracting web requests specifically, a wealth of work exists for analyzing the strings that can appear in an application -- similar to what \stringoid does. 
Automata have been used before to represent strings \cite{Christensen:2003}; a key issue in general-purpose string analysis is whether widening is used to make the analysis more tractable as in~\cite{Yu08}. In \stringoid we have side-stepped the issue of loops by focusing on straight-line constructions, which we believe cover the majority of URLs.
Other abstract representations for string values have been tried; the HAMPI~\cite{Kiezun:2009:HSS:1572272.1572286} solver for instance builds a customized efficient representation, which can be used to efficiently check the presence of certain substrings, but it requires a finite bound on strings. Because we are focused on enumerating possible strings rather than checking them against a candidate language as is common in security applications, and because we did not want to a priori impose length constraints, we favored automata.

Static analyses are typically evaluated with respect to a known ground truth. Our work in contrast compares the results of \stringoid to data obtained dynamically, providing a unique perspective on the strengths and limitations of both approaches in a concrete use case.


Android applications have been mined, among other things, for security credentials~\cite{viennot2014measurement},
for automatically generating graphical user-interface test cases~\cite{Linares-Vasquez:2015}, 
to understand how power consumption of mobile devices is managed~\cite{Bao:2016}, or to detect malware~\cite{Yang2014} or data leaks~\cite{Arzt:2014}. However, to the best of our knowledge, Android applications have not yet been mined at scale to detect web requests.

Beyond the here used Playdrone dataset of Android applications~\cite{viennot2014measurement}, other datasets have been published, that provide source code version histories of selected Android applications~\cite{Krutz:2015}, or follow a continuous mining strategy to reflect the ever-growing Android ecosystem~\cite{Allix:2016}. We chose Playdrone when starting our work as the only dataset that made available the desired amount of application binaries.


\section{Conclusion}
\label{sec:conclusion}

We explored the potential for analyzing web requests made by Android
applications in two ways; our dynamic analysis, through instrumentation,
reveals common communication patterns.  Thanks to the use of an HTTPS proxy, we
were able to produce a dataset of complete sets of requests and responses for
20 applications exercised manually.  Furthermore, we manually classified all
requests according to the likely origin of their URL. Ultimately, we
demonstrated that a large number of web requests made by applications are not
immediately traceable to their code. 

In contrast, static analysis, our second approach, while carefully designed, is
a blunt tool; our analysis can extract large amounts of URL components from a
large number of application with no supervision, making it amenable to large-scale
studies. We have hinted at some of the possibilities opened up by our
dataset and our tool \stringoid. On applications for which we could compare both
approaches, we discovered, through the results of static analysis, URL and URL
components which we were not able to exercise dynamically despite our best
efforts. While we cannot conclusively establish that these are valid and live
in the application, it demonstrates the potential of static analysis tools such
as \stringoid to assist in coverage, and to map out APIs from unknown
applications.


\section*{Acknowledgments}

We thank the anonymous reviewers for their thoughtful comments, and Alex Alekseyev for his help and comments.

\newpage
\balance
\bibliographystyle{IEEEtran}
\bibliography{bibliography}
%



\end{document}